\documentclass[conference, onecolumn]{IEEEtran}
\usepackage{epsfig,times,graphics}
\usepackage{amsmath,amsfonts,amssymb}
\usepackage{cite,graphicx,subfigure,url}
\usepackage{setspace}

\newcommand{\SF}{ {\mathcal F} }

\newcommand {\tE}{\textrm{E}}

\newtheorem{definition}{{\bf Definition}}
\newtheorem{theorem}{{\bf Theorem}}

\newtheorem{proposition}{\hspace{-0.15in}{\bf Proposition}}

\def\done{\hspace*{\fill} \rule{1.8mm}{2.5mm} }

\begin{document}

\title{{\bf Applying Stochastic Network Calculus to 802.11 Backlog and Delay Analysis}}

\author{Yue Wang\\
        Department of Computer Science \& Engineering\\
        The Chinese University of Hong Kong \\
        Email: ywang@cse.cuhk.edu.hk
}

\maketitle

\begin{abstract}
Stochastic network calculus provides an elegant way to characterize
traffic and service processes. However, little effort has been made
on applying it to multi-access communication systems such as 802.11.
In this paper, we take the first step to apply it to the backlog and
delay analysis of an 802.11 wireless local network. In particular,
we address the following questions: In applying stochastic network
calculus, under what situations can we derive stable backlog and
delay bounds? How to derive the backlog and delay bounds of an
802.11 wireless node? And how tight are these bounds when compared
with simulations? To answer these questions, we first derive the
general stability condition of a wireless node (not restricted to
802.11). From this, we give the specific stability condition of an
802.11 wireless node. Then we derive the backlog and delay bounds of
an 802.11 node based on an existing model of 802.11. We observe that
the derived bounds are loose when compared with ns-2 simulations,
indicating that improvements are needed in the current version of
stochastic network calculus.
\end{abstract}

\section{{\bf Introduction}}\label{sec:introduction}

Network calculus provides an elegant way to characterize traffic and
service processes of network and communication systems. Unlike
traditional queueing analysis in which one has to make strong
assumptions on arrival or service processes (e.g., Poission arrival
process, exponential service distribution, etc) so as to derive
closed-form solutions\cite{queueing_book}, network calculus allows
general arrival and service processes. Instead of getting exact
solutions, one can derive network delay and backlog bounds easily by
network calculus. Deterministic network calculus was proposed in
\cite{netcal_cruz_1}
\cite{netcal_cruz_2}\cite{netcal_boudec}\cite{netcal_chang}, etc.
However, most traffic and service processes are stochastic and
deterministic network calculus is often not applicable for them.
Therefore, stochastic network calculus was proposed to deal with
stochastic arrival and service processes
\cite{netcal_chang}\cite{snetcal_cruz}\cite{snetcal_liu}\cite{snetcal_jiang_o1}\cite{snetcal_jiang_o2}\cite{snetcal_jiang}\cite{snetcal_book_jiang}.

There have been some applications of stochastic network
calculus\cite{app_MBAC}\cite{app_SLA}\cite{app_LRD_GPS}\cite{app_wireless}.
However, little effort has been made on applying it to multi-access
communication systems. In the paper, we take the first step to apply
stochastic network calculus to an 802.11 wireless local network
(WLAN). In particular, we address the following questions:
\begin{itemize}
\item Under what situations can we derive stable backlog and delay bounds?
\item How to derive the backlog and delay bounds of an 802.11 wireless node?
\item How tight are these bounds when compared with simulations?
\end{itemize}

In this paper, we answer these questions and make the following
contributions:
\begin{itemize}
\item We derive the general stability condition of a wireless node based on
the theorems of stochastic network calculus. From this, we give the
specific stability condition of an 802.11 wireless node.
\item We derive the service curve of an
802.11 node based on an existing model of 802.11\cite{Kumar_802_11}.
From the service curve, we then derive the backlog and delay bounds
of the node.
\item The derived bounds are loose in many cases
when compared with ns-2 simulations. We discuss the reasons and
point out future work.
\end{itemize}

This paper is organized as follows. In Section~\ref{sec:snc}, we
give a brief overview of stochastic network calculus. In
Section~\ref{sec:model}, we present the stochastic network calculus
model of a wireless node. In Section~\ref{sec:stab}, we derive the
general stability condition of a wireless node. In
Section~\ref{sec:802_11}, we derive the backlog and delay bounds and
the stability condition for an 802.11 node. In
Section~\ref{sec:simulation}, we compare the derived bounds with
simulation results. Related work is given in
Section~\ref{sec:related} and finally, Section~\ref{sec:conclusion}
concludes the paper and points out future directions.

\section{{\bf Stochastic Network Calculus}}\label{sec:snc}

In this section, we first review basic terms of network calculus and
then cite the results of stochastic network calculus which we will
use in this paper. There are various versions of arrival and service
curves. We adopt \emph{virtual backlog centric (v.b.c) stochastic
arrival curve} and \emph{weak stochastic service curve} in our
analysis.

\subsection{{\bf Basic Terms of Network Calculus}}

\begin{figure}[htb]
  \centering
  \includegraphics[width=0.8\textwidth]{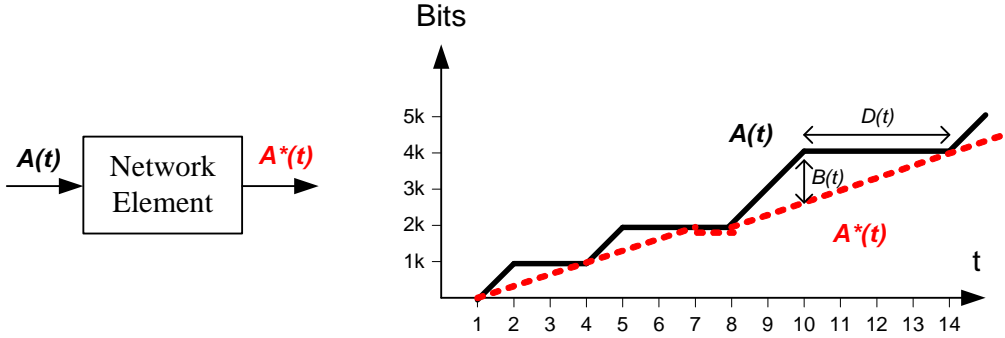}\\
  \caption{Illustration of $A(t)$, $A^*(t)$, $B(t)$ and $D(t)$}\label{fig:curves_eg}
\end{figure}

We consider a discrete time system where time is slotted ($t = 0, 1,
2, ...$). A process is a function of time $t$. By default, we use
$A(t)$ to denote the \emph{arrival process} to a network element
with $A(0)=0$. $A(t)$ is the total amount of traffic arrived to this
network element up to time $t$. We use $A^*(t)$ to denote the
\emph{departure process} of the network element with $A^*(0)=0$.
$A^*(t)$ is the total amount of traffic departed from the network
element up to time $t$. Let $\SF$ ($\bar{\SF}$) represents the set
of non-negative wide-sense increasing (decreasing) functions.
Clearly, $A(t)\in \SF$ and $A^*(t)\in \SF$. For any process, say
$A(t)$, we define $A(s,t) \equiv A(t)-A(s)$, for $s \leq t$. We
define the backlog of the network element at time $t$ by
\begin{eqnarray}\label{eq:backlog}
B(t)=A(t)-A^*(t),
\end{eqnarray}
and the delay of the network element at $t$ by
\begin{eqnarray}\label{eq:delay}
D(t)=\inf\{\tau: A(t)\leq A^*(t+\tau)\}.
\end{eqnarray}
Fig.~\ref{fig:curves_eg} illustrates an example of $A(t)$ and
$A^*(t)$ with $B(t)$ and $D(t)$ at $t=10$.

In deterministic network calculus, $A(t)$ can be upper-bounded by an
arrival curve. That is, for all $0 \leq s \leq t$, we have
\[A(s, t) \leq \alpha(t-s),\]
where $\alpha(t)$ is called the \emph{arrival curve} of $A(t)$.

A \emph{busy period} is a time period during which the backlog in
the network element is always nonzero. For any busy period $(t_0,
t]$, suppose we have
\[A^*(t) - A^*(t_0) \geq \beta(t-t_0),\]
which means that the network element provides a guaranteed service
lower-bounded by $\beta(t-t_0)$ during the busy period. We can let
$t_0$ be the beginning of the busy period, that is, the backlog at
$t_0$ is zero or $A^*(t_0)=A(t_0)$. Therefore,
\[A^*(t) - A(t_0) \geq \beta(t-t_0).\]

The above equation infers $A^*(t) \geq \inf_{0\leq s \leq
t}{[A(s)+\beta(t-s)]}$, which can be written as
\begin{eqnarray}
A^*(t) \geq A\otimes\beta(t),
\end{eqnarray}
where $\otimes$ is called the operator of \emph{min-plus
convolution} and $\beta(t)$ is called the \emph{service curve} of
the network element.

\subsection{{\bf Results of Stochastic Network Calculus}}

We cite the following definitions and theorems from
\cite{netcal_chang}\cite{snetcal_jiang} except that we define
Definition~\ref{def:stability} by ourselves.

\begin{definition}[virtual-backlog-centric (v.b.c)
Stochastic Arrival Curve]\label{def:ac} A flow is said to have a
virtual-backlog-centric (v.b.c) stochastic arrival curve $\alpha \in
\SF$ with bounding function $f \in \bar{\SF}$, denoted by $A
\sim_{vb} <f, \alpha>$, if for all $t \geq 0$ and all $x \geq 0$,
there holds
\begin{eqnarray}
P\{\sup_{0 \leq s \leq t}[A(s, t) - \alpha(t-s)]>x\} \leq f(x).
\end{eqnarray}
\end{definition}

Originally, in deterministic network calculus, we have $A(s, t) \leq
\alpha(t-s)$ for all $0\leq s \leq t$. However, there is usually
some randomness in stochastic arrival processes and $A(s,t)$ may not
be upper-bounded by any arrival curve deterministically (e.g.,
traffic arrivals in $(s, t]$ can be arbitrarily large in Poisson
process). Thus, v.b.c stochastic arrival curve is proposed to tackle
this problem. Roughly speaking, $A(s,t)$ can exceed $\alpha(s,t)$ by
$x$, but its probability is upper-bounded by $f(x)$ which is a
decreasing function of $x$.

\begin{definition}[Weak Stochastic Service Curve]\label{def:sc}
A server $S$ is said to provide a weak stochastic service curve
$\beta \in \SF$ with bounding function $g \in \bar{\SF}$, denoted by
$S \sim_{ws} <g, \beta>$, if for all $t \geq 0$ and all $x \geq 0$,
there holds
\begin{eqnarray}
P\{A \otimes \beta(t) - A^*(t) > x\} \leq g(x).
\end{eqnarray}
\end{definition}

In deterministic network calculus, we have $A^*(t) \geq A \otimes
\beta(t)$, which means that there is a service guarantee denoted by
the service curve $\beta(t)$. However, there is usually some
randomness in stochastic service process and thus a server may not
always provide a guaranteed service curve deterministically. Thus,
weak stochastic service curve is proposed to tackle this problem.
Roughly speaking, $A^*(t) - A \otimes \beta(t)$ can be less than
$-x$, but its probability is upper-bounded by $g(x)$ which is a
decreasing function of $x$.

The utility of the above definitions is that if we can characterize
the traffic by a v.b.c stochastic arrival curve and the server's
service process by a weak stochastic service curve, then we can
calculate backlog and delay bounds of the network element by
Theorem~\ref{theo:backlog_delay}.

\begin{theorem}[Backlog and Delay Bounds]\label{theo:backlog_delay}
Consider a server fed with a flow $A$. If the server provides a weak
stochastic service curve $S \sim_{ws} <g, \beta>$ to the flow and
the flow has a v.b.c stochastic arrival curve $A \sim_{vb} <f,
\alpha>$, then

(i) The backlog $B(t)$ of the flow in the server at time $t$
satisfies: for all $t \geq 0$ and all $x \geq 0$,
\begin{eqnarray}
P\{B(t) > x\} \leq f \otimes g(x+\inf_{s \geq
0}{[\beta(s)-\alpha(s)]}).
\end{eqnarray}

(ii) The delay $D(t)$ of the flow in the server at time $t$
satisfies: for all $t \geq 0$ and all $x \geq 0$,
\begin{eqnarray}
P\{D(t) > x\} \leq f \otimes g(\inf_{s \geq
0}{[\beta(s)-\alpha(s-x)]}).
\end{eqnarray}
\end{theorem}
By definition $\alpha(x)=0$ when $x<0$ in this theorem. Note that as
noticed recently by researchers of network calculus, the formula of
delay bound in this theorem often returns trivial results, which we
will see in Section~\ref{sec:802_11}.

By now, we have reviewed the key results of stochastic network
calculus. Next, we will show how to calculate v.b.c stochastic
arrival curve and weak stochastic service curve.

In \cite{snetcal_jiang}, the author presented a theorem to
facilitate calculation of stochastic arrival curves. Before showing
the theorem, we first introduce $(\sigma(\theta),
\rho(\theta))$\emph{-upper constrained} \cite{netcal_chang}.

\begin{definition}[$(\sigma(\theta), \rho(\theta))$-upper constrained]\label{def:theta_uc}
A process $A$ is said to be $(\sigma(\theta), \rho(\theta))$-upper
constrained (for some $\theta > 0$), if for all $0 \leq s \leq t$,
we have
\begin{eqnarray}
\frac{1}{\theta} \log{\tE e^{\theta A(s,t)}} \leq \rho(\theta)(t-s)
+ \sigma(\theta).
\end{eqnarray}
\end{definition}

This definition is equivalent to $\tE e^{\theta A(s,t)} \leq
e^{\theta \rho(\theta)(t-s) + \theta \sigma(\theta)}$, which means
that $A(s,t)$'s moment generating function is upper-bounded. Two
related concepts are defined as follows.

\begin{definition}[$\theta$-MER / $\theta$-ER]\label{def:theta-MER}
A process $A$'s \emph{minimum envelope rate (MER) with respect to
$\theta$} ($\theta$-MER), denoted by $\rho^*(\theta)$, is defined as
follows:
\begin{eqnarray}
\rho^*(\theta) = \lim\sup_{t\rightarrow \infty}\frac{1}{\theta
t}\sup_{s\geq 0}\log{\tE e^{\theta A(s,s+t)}}.
\end{eqnarray}
We say that $A$ has an \emph{envelope rate (ER) with respect to
$\theta$} ($\theta$-ER), denoted by $\rho(\theta)$,  if
$\rho(\theta) \geq \rho^*(\theta)$.
\end{definition}

The following theorem expresses the relationship between $\theta$-ER
and $(\sigma(\theta), \rho(\theta))$-upper constrained.

\begin{theorem}[Relationship of
$\theta$-ER and $(\sigma(\theta), \rho(\theta))$-upper
constrained]\label{theo:relation_theta}\ \\ (i) If $A$ is
$(\sigma(\theta), \rho(\theta))$-upper constrained, then
$\rho(\theta)$ is $\theta$-ER of $A$. \ \\(ii) If $A$ has
$\theta$-ER $\rho(\theta)<\infty$, then for every $\epsilon > 0$
there exists $\sigma_\epsilon(\theta) < \infty$ so that $A$ is
$(\sigma_\epsilon(\theta), \rho(\theta)+\epsilon)$-upper
constrained.
\end{theorem}


Now we have two kinds of traffic characterization: v.b.c stochastic
arrival curve and $(\sigma(\theta), \rho(\theta))$-upper
constrained. The following theorem establishes the connection
between them.

\begin{theorem}[v.b.c Stochastic Arrival Curve of
$(\sigma(\theta), \rho(\theta))$-upper constrained
Process]\label{theo:ac_theta} Suppose $A(t)$ is
$(\sigma(\theta),\rho(\theta))$-upper constrained, then it has a
v.b.c stochastic arrival curve\footnote{The original theorem
(Theorem~5.1\cite{snetcal_jiang}) established a similar connection
between maximum-backlog-centric (m.b.c) stochastic arrival curve and
$(\sigma(\theta), \rho(\theta))$-upper constrained, which is wrong
as noticed recently by researchers of network calculus. However, one
can easily see the theorem holds for v.b.c stochastic arrival
curve.} $A \sim_{vb} <f, \alpha>$, where
\begin{eqnarray}\label{eq:theo_ac_theta}
\alpha(t) &=& r \cdot t\nonumber\\
f(x) &=&
\frac{e^{\theta\sigma(\theta)}}{1-e^{\theta(\rho(\theta)-r)}}\cdot
e^{-\theta x}
\end{eqnarray}
for any $r > \rho(\theta)$ and $x \geq 0$.
\end{theorem}

This theorem indicates that if we can show that the traffic is
$(\sigma(\theta), \rho(\theta))$-upper constrained, then we can get
its v.b.c stochastic arrival curve by Eq.~(\ref{eq:theo_ac_theta}).

We now introduce the concept of \emph{stochastic strict server}.
This concept was inspired by the observation that a wireless channel
can be described by an ideal service process and an impairment
process. As we will see in Section~\ref{sec:model}, a wireless node
can be modeled as a stochastic strict server.

\begin{definition}[Stochastic Strict Server]
A server $S$ is said to be a stochastic strict server providing
stochastic strict service curve $\hat{\beta} \in \SF$ with
impairment process $I$ to a flow iff during any backlogged period
$(s, t]$, the output $A^*(s, t)$ of the flow from the server
satisfies
\begin{eqnarray}
A^*(s,t) \geq \hat{\beta}(t-s)-I(s,t).
\end{eqnarray}
\end{definition}

We can easily find the weak stochastic service curve of a stochastic
strict server by the following theorem.
\begin{theorem}[Weak Stochastic Service Curve of Stochastic Strict Server]
\label{theo:strict_sc} Consider a stochastic strict server $S$
providing a stochastic strict service curve $\hat{\beta}$ with an
impairment process $I$. If the impairment process has a v.b.c
stochastic arrival curve, or $I \sim_{vb} <g, \gamma>$, and $\beta
\in \SF$, then the server provides a weak stochastic service
curve\footnote{In the original theorem
(Lemma~4.2\cite{snetcal_jiang}), the impairment process has a m.b.c
stochastic arrival curve. However, one can easily see that the
theorem also holds for the impairment process which has a v.b.c
stochastic arrival curve.} $S \sim_{ws} <g, \beta>$ with
\begin{eqnarray}
\beta(t) = \hat{\beta}(t) - \gamma(t).
\end{eqnarray}
\end{theorem}

So far, we have cited all results of stochastic network calculus
which we will use in this paper. Finally, we define stable backlog
and stable delay. A natural definition is to check whether the
expectation of backlog (or delay) is finite.

\begin{definition}[Stable Backlog/Delay]\label{def:stability}
The backlog $B(t)$ is stable, if
\begin{eqnarray}
\tE B(t) < \infty,  \quad \forall t.
\end{eqnarray}
Similarly, the delay $D(t)$ is stable, if
\begin{eqnarray}
\tE D(t) < \infty,  \quad \forall t.
\end{eqnarray}
\end{definition}

We say that the backlog (or delay) bound of stochastic network
calculus is stable if they can derive stable backlog (or delay).

\section{{\bf Stochastic Network Calculus Model of a Wireless Node}}\label{sec:model}
In this section we model a wireless node (not restricted to 802.11)
by stochastic network calculus. In general, we can define one slot
($t=1$) to be any duration of time and measure traffic amount in any
unit (e.g. bits, bytes or packets).

We consider a wireless node. Let $A(t)$ denote the traffic arrived
at the node from the application layer. Suppose $A$ is
$(\sigma_{A}(\theta_1), \rho_{A}(\theta_1))$-upper constrained. From
Theorem~\ref{theo:ac_theta}, we have $A \sim_{vb} <f,\alpha>$, where
\begin{eqnarray}\label{eq:A_ac}
\alpha(t) &=& r_A \cdot t\nonumber\\
f(x) &=&
\frac{e^{\theta_1\sigma_{A}(\theta_1)}}{1-e^{\theta_1(\rho_{A}(\theta_1)-r_A)}}\cdot
e^{-\theta_1 x}
\end{eqnarray}
for any $r_A > \rho_{A}(\theta_1)$.

We can model a wireless node by a stochastic strict server. Let the
channel capacity be $c$ traffic unit per slot. The departure process
$A^*(s, t) = \hat{\beta}(s, t) - I(s, t)$ during any backlogged
period $(s,t]$, where $\hat{\beta}(t) = c\cdot t$ is the ideal
service curve and $I$ is the impairment process due to backoff,
channel sharing and transmission errors. Since $I(s,t) \leq
c\cdot(t-s)$, $I$ has a finite $\theta$-MER. From
Theorem~\ref{theo:relation_theta}, there exist
$\sigma_{I}(\theta_2)$ and $\rho_{I}(\theta_2)$ so that $I$ is
$(\sigma_{I}(\theta_2), \rho_{I}(\theta_2))$-upper constrained.
Based on Theorem~\ref{theo:ac_theta}, we have $I \sim_{vb}
<g,\gamma>$, where
\begin{eqnarray}\label{eq:I_ac}
\gamma(t) &=& r_I \cdot t\nonumber\\
g(x) &=&
\frac{e^{\theta_2\sigma_{I}(\theta_2)}}{1-e^{\theta_2(\rho_{I}(\theta_2)-r_I)}}\cdot
e^{-\theta_2 x},
\end{eqnarray}
for any $r_I > \rho_{I}(\theta_2)$.

From Theorem~\ref{theo:strict_sc}, the node provides a weak
stochastic service curve $S \sim_{ws} <g, \beta>$, where
\begin{eqnarray}\label{eq:S_sc}
\beta(t) &=& (c-r_I) \cdot t,
\end{eqnarray}
for any $c > r_I$.

Furthermore, from Theorem~\ref{theo:backlog_delay}, we must have
$\alpha(t) \leq \beta(t)$, or equivalently,
\begin{eqnarray}\label{eq:stab_1}
r_A \leq c - r_I.
\end{eqnarray}
Thus, $P\{B(t)>x\} \leq f\otimes g(x)$. Otherwise, if $\alpha(t) >
\beta(t)$, we get a trivial backlog bound, $P\{B(t)>x\} \leq
f\otimes g(-\infty) = \infty$.

\section{{\bf Stability Condition of a Wireless Node}}\label{sec:stab}

One fundamental question we need to address is under what condition
we can get \emph{stable} $B(t)$ and $D(t)$ from stochastic network
calculus, i.e., $\tE B(t) < \infty$ and $\tE D(t) < \infty$. Before
presenting our result, we first define the concept of \emph{envelop
average rate}.

\begin{definition}[Envelop Average Rate]\label{def:ER} The envelop average rate of a process
$A$, denoted by $a_A$, is defined as
\begin{eqnarray}
a_A = \lim_{t\rightarrow \infty}\sup_{s\geq 0}\frac{\tE
A(s,s+t)}{t}.
\end{eqnarray}
\end{definition}

Let $a_{A}$ and $a_{I}$ be the envelop average rate of $A$ and $I$,
respectively. The following proposition shows the stability
condition.

\begin{proposition}[Stability Condition]\label{prop:stab}
A wireless node has stable backlog and stable delay if
\begin{eqnarray}\label{eq:stab}
a_{A} < c-a_{I}.
\end{eqnarray}
\end{proposition}

\noindent\emph{Proof:} We have shown that $P\{B(t)>x\} \leq f\otimes
g(x)$ if Eq.~(\ref{eq:stab_1}) holds. Thus, for any $t$,
\begin{eqnarray}\label{eq:EB}
\tE B(t) &=& \sum_{i=0}^\infty P\{B(t) = i+1\} \cdot (i+1)\nonumber\\
&<& \sum_{i=0}^\infty P\{B(t) > i \} \cdot (i+1)\nonumber\\
&\leq& \sum_{i=0}^\infty f\otimes g(i) \cdot (i + 1) < \infty.
\end{eqnarray}

Since $f(x)$ and $g(x)$ are exponentially decreasing functions
according to Eq.~(\ref{eq:A_ac}) and Eq.~(\ref{eq:I_ac}), $f\otimes
g(x)$ is an exponentially decreasing function. Thus we have, for any
$t$, $\tE B(t) < \infty$. It is easy to see that for any $t$, $\tE
D(t) < \infty$. Otherwise, the service time is $\infty$ and thus
$\lim_{t\rightarrow\infty} \tE B(t) = \infty$ which contradicts
Eq.~(\ref{eq:EB}).

Now we examine Eq.~(\ref{eq:stab_1}). From Eq.~(\ref{eq:A_ac}) and
Eq.~(\ref{eq:I_ac}), $r_A = \rho_{A}(\theta_1) + \epsilon$ and $r_I
= \rho_{I}(\theta_2) + \epsilon$ for any $\epsilon >0$. Thus,
Eq.~(\ref{eq:stab_1}) is equivalent to
\begin{eqnarray}\label{eq:stab_3}
\rho_{A}(\theta_1) \leq c - \rho_{I}(\theta_2) - 2\epsilon.\nonumber
\end{eqnarray}

From Theorem~\ref{theo:relation_theta}, we have
$\rho_{A}(\theta_1)=\rho_{A}^*(\theta_1)+\epsilon_1$ and
$\rho_{I}(\theta_2)=\rho_{I}^*(\theta_2)+\epsilon_1$ for any
$\epsilon_1
> 0$, where $\rho_{A}^*(\theta_1)$ and $\rho_{I}^*(\theta_2)$ are $\theta$-MERs of $A$ and $I$, respectively.
Equivalently, we have
\begin{eqnarray}\label{eq:stab_4}
\rho_{A}^*(\theta_1) \leq c-\rho_{I}^*(\theta_2) -
2(\epsilon+\epsilon_1).
\end{eqnarray}

Using Taylor expansion on $\rho_{A}^*(\theta_1)$, we have
\begin{eqnarray*}
\rho_{A}^*(\theta_1) &=& \lim\sup_{t\rightarrow
\infty}\frac{1}{\theta_1
t}\sup_{s\geq 0}\log{\tE e^{\theta_1 A(s,s+t)}}\\
&=& \lim {\sup_{t\rightarrow \infty}{\frac{1}{\theta_1
t}\sup_{s\geq 0}{\log{\tE (1 + \theta_1 A(s,s+t) + O(\theta_1^2))}}}}\\
&=& \lim{\sup_{t\rightarrow \infty}{\frac{1}{\theta_1
t}\sup_{s\geq 0}{\log{(1 + \theta_1\tE A(s,s+t) + O(\theta_1^2))}}}}\\
&=& \lim{\sup_{t\rightarrow \infty}{\frac{1}{\theta_1 t}\sup_{s\geq
0}{[\theta_1\tE A(s,s+t) + O(\theta_1^2)]}}}.
\end{eqnarray*}

Therefore,
\begin{eqnarray}
\lim_{\theta_1\rightarrow 0}\rho_{A}^*(\theta_1) =
\lim_{t\rightarrow \infty}\sup_{s\geq 0}{\frac{\tE A(s,s+t)}{t}} =
a_{A}.\nonumber
\end{eqnarray}

Similarly,
\begin{eqnarray}
\lim_{\theta_2\rightarrow 0}\rho_{I}^*(\theta_2) = a_{I}.\nonumber
\end{eqnarray}

Therefore, there exist $\theta_1$ and $\theta_2$ so that
$\rho_{A}^*(\theta_1)\leq a_{A}+\epsilon_2$ and
$\rho_{I}^*(\theta_2)\leq a_{I}+\epsilon_2$ for any enough small
$\epsilon_2 > 0$. So Eq.~(\ref{eq:stab_4}) is satisfied if
\begin{eqnarray}\label{eq:stab_5}
a_{A} \leq c-a_{I} - 2(\epsilon+\epsilon_1+\epsilon_2).\nonumber
\end{eqnarray}
Since $\epsilon$, $\epsilon_1$ and $\epsilon_2$ can be arbitrarily
small, the above equation is satisfied when
\begin{eqnarray}\label{eq:stab_6}
a_{A} < c-a_{I}.\nonumber
\end{eqnarray}
\done
\\
\noindent\textbf{Remarks:} Since the proof is based on theorems of
stochastic network calculus, it indicates that we can get stable
backlog and delay bounds by stochastic network calculus as long as
the condition of Eq.~(\ref{eq:stab}) holds. Since this condition is
very general, we conclude that theoretically stochastic network
calculus is effective.

\section{{\bf 802.11 Backlog and Delay Bounds}}\label{sec:802_11}

In this section, we apply the results in the previous section to
calculate backlog and delay bounds for an 802.11 WLAN node. For
simplicity, we assume there are $n$ \emph{identical} stations (or
nodes) sending packets to an access point. All nodes operate in
Distributed Coordination Function (DCF) mode with RTS/CTS turned
off\cite{802_11}. We consider an ideal channel, that is,
transmission errors are only caused by collisions. Two packets are
collided if their transmissions overlap in time.
Besides, we assume that all DATA packets are of the same size.

\subsection{{\bf 802.11 DCF}}\label{sec:802_11_DCF}
A node with a DATA packet (or simply packet) to transmit first
monitors the channel activity. If the channel is idle for a period
of time equal to a distributed interframe space (DIFS), the node
transmits. Otherwise, if the channel is sensed busy (either
immediately or during the DIFS), the node backs off, in which the
node defers channel access by a random number of \emph{idle slots}
within a contention window ($CW$), ranging from 0 to $CW-1$. When
the backoff counter reaches zero and expires, the node can access
the channel. During the backoff period, if the node detects a busy
channel, it freezes the backoff counter and the backoff process is
resumed once the channel is idle for a duration of DIFS. To avoid
channel capture, a node must wait a random backoff time between two
consecutive new packet transmissions, even if the channel is sensed
idle for a duration of DIFS. Once the packet is received
successfully, the receiver will return an ACK after a short
interframe space (SIFS). Note that SIFS is shorter than an idle slot
so that there is no collision caused by a DATA packet and an ACK.

802.11 uses the truncated exponential backoff technique to set its
$CW$. For example, in 802.11b, the initial $CW$ is $CW_{min}=32$.
Each time a collision occurs, $CW$ doubles its size, up to a maximum
of $CW_{max}=1024$. When the packet is successfully transmitted,
$CW$ is reset to $CW_{min}$. The packet is dropped when it is
retransmitted for \emph{six} times and still not transmitted
successfully. Fig.~\ref{fig:802_11} shows some parameters of 802.11b
used in our paper.

\begin{figure}[htb]
\begin{center}
\begin{tabular}{|l|l|}
\hline
Basic rate & 1 Mbps\\
Data rate & 11 Mbps\\
PHY header & 24 bytes\\
ACK header & 14 bytes\\
MAC header & 28 bytes\\
SIFS & 10 $\mu$s\\
DIFS & 50 $\mu$s\\
Idle slot & 20 $\mu$s\\
$CW_{min}$ & 32\\
$CW_{max}$ & 1024\\
Retransmission limit & 6\\
\hline
\end{tabular}
\end{center}
\caption{802.11b parameters}\label{fig:802_11}
\end{figure}

The duration of an ACK is the duration of a PHY header and an ACK
header transmitted at \emph{basic rate}, i.e., $\frac{(24+14)\cdot
8}{10^6} = 304\mu$s. The duration of a DATA packet is the duration
of an PHY header transmitted at \emph{basic rate} plus the duration
of an MAC header and its upper-layer payload transmitted at
\emph{data rate}. For example, suppose the upper-layer payload is
$256$ bytes, then the duration of the DATA packet is $\frac{24\cdot
8}{10^6}+\frac{(28+256) \cdot 8}{11\cdot 10^6}=398.5\mu$s.


\subsection{{\bf Service Curve}}\label{sec:802_11_sc}
Since we only consider packets of equal size, we can measure traffic
amount in packets. For simplicity, we measure time duration (e.g.
SIFS, DIFS, DATA and ACK) in idle slots and we define that one slot
in network calculus ($t=1$) is equal to $L$ idle slots, where
\begin{eqnarray}\label{eq:L}
L = DIFS+DATA+SIFS+ACK.
\end{eqnarray}
Note that sometimes "idle slot" only refers to a time period and it
may not be idle. To avoid confusion in the following context, we
will use "\emph{idle slot}" (italic) to denote that the "idle slot"
is indeed idle.

In practice, it is difficult to calculate the impairment process $I$
accurately since $I$ depends on the complex interactions of traffic
arrival and DCF. In this section, we perform the worst case analysis
based on an existing model of 802.11\cite{Kumar_802_11}.

We assume that the system is in saturated condition, that is, the
backlog at each node is always nonzero. Let $\tau$ denote the
transmission attempt probability per \emph{idle slot} by a node and
let $\gamma$ denote the conditional collision probability given that
there is a transmission. We assume $\gamma$ is constant and
independent for each transmission. Intuitively, this assumption
becomes more accurate when the number of nodes $n$ increases. In
\cite{Kumar_802_11}, the authors derived a general formula relating
$\tau$ to $\gamma$, which is
\begin{eqnarray}\label{eq:tau_gamma}
\tau = \frac{1+\gamma+\gamma^2+...+\gamma^6}{b_0+\gamma b_1+\gamma^2
b_2+...+\gamma^6 b_6}.
\end{eqnarray}
This equation can be explained as follows. The numerator is the
expected number of transmission attempts of a packet. The
denominator is the expected total backoff duration (in \emph{idle
slots}) of a packet, where $b_i$ is the mean backoff duration plus 1
(the 1 refers to the first idle slot of packet transmission) after
the $i$th collision. In 802.11, $b_i = \frac{2^i \cdot CW_{min}}{2}$
where $0 \leq i \leq 6$. A packet suffering six collisions will be
dropped from its buffer. In our model, we do not consider packet
drops, which yields upper bounds of backlog and delay.

The independence assumption of $\gamma$ implies that each
transmission "sees" the system at steady state. Therefore, each node
transmits with the same probability $\tau$. This yields
\begin{eqnarray}\label{eq:gamma_tau}
\gamma = 1-(1-\tau)^{n-1}.
\end{eqnarray}
Combining Eq.~(\ref{eq:tau_gamma}) and Eq.~(\ref{eq:gamma_tau}), we
can solve $\tau$ and $\gamma$.

We introduce the following notations for an 802.11 node. The
probability of no transmissions at an \emph{idle slot}, denoted by
$P_{nt}$, is $(1-\tau)^n$. The probability of having at least one
transmission at an \emph{idle slot}, denoted by $P_t$, is
$1-P_{nt}$. The probability of a \emph{given} node starting a
successful transmission at an \emph{idle slot}, denoted by $P_s$, is
$\tau (1-\gamma)$. For the given node, the probability of the other
nodes starting transmissions at an \emph{idle slot}, denoted by
$P_o$, is $P_t - P_s = \gamma$.

\begin{figure}[htb]
  \centering
  \includegraphics[width=0.6\textwidth]{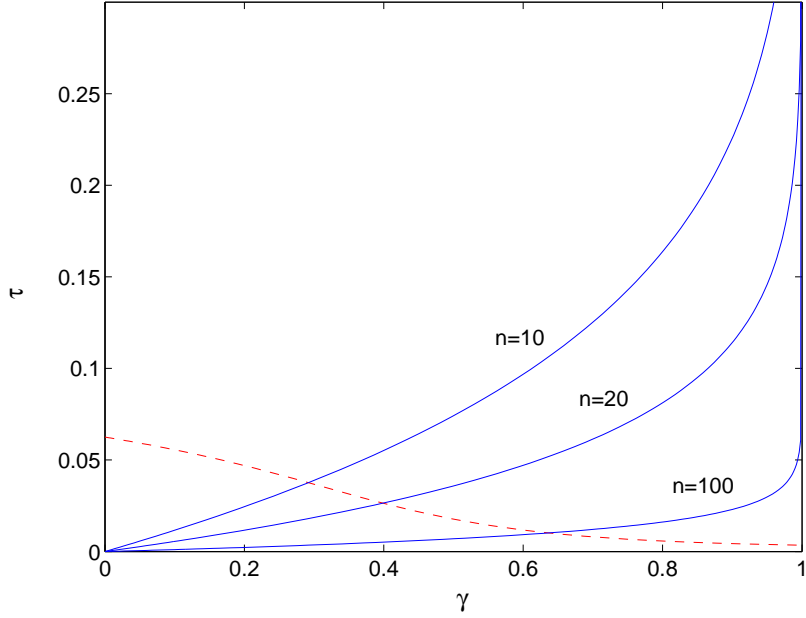}\\
  \caption{The Plots of Eq.~(\ref{eq:tau_gamma}) and Eq.~(\ref{eq:gamma_tau})}\label{fig:gamma_tau}
\end{figure}

Fig.~\ref{fig:gamma_tau} plots Eq.~(\ref{eq:tau_gamma}) in dashed
line and Eq.~(\ref{eq:gamma_tau}) in solid line when $n = 10$, $20$
and $100$. The points of intersection are the solution of $\gamma$
and $\tau$ for different $n$. When $n$ increases, $\gamma$ increases
and $\tau$ decreases. Consequently, $P_{nt}$ and $P_o$ increases,
but $P_s$ decreases. Therefore, the assumption of saturated
condition gives the \emph{worst-case} analysis.

An 802.11 node can be seen as a stochastic strict server. Clearly,
the stochastic strict service curve $\hat{\beta}(t)=t$, which means
that one packet is transmitted during one slot in the ideal case.
In order to characterize the impairment process $I$, it is crucial
to know $\tE e^{\theta I(s,s+t)}$ which we calculate as follows.

We consider the transmissions of a given 802.11 node during $t$
slots. From Eq.~(\ref{eq:L}), there are $tL$ idle slots in $t$
slots, indexed $1$, $2$, ..., $tL$. At the first or the last idle
slot, the transmission (if any) can be \emph{incomplete}, that is,
the transmission can start before the first idle slot or it can end
after the last idle slot. We assume that at the first slot there is
always a \emph{complete} transmission by any of the other nodes
except the given node, which actually overestimates $\tE e^{\theta
I(s,s+t)}$. Suppose there are $i$ \emph{complete} transmissions and
zero or one \emph{incomplete} transmission within the remaining
$(t-1)L$ idle slots. The incomplete transmission occupies the last
$k$ idle slots where $0 \leq k \leq L-1$ ($k=0$ means that the last
transmission is actually complete). Thus, there are $(t-i-1)L-k$
\emph{idle slots} of no transmissions within the $t$ slots. The
probability of $i$ complete transmissions and the last $k$ idle
slots occupied by an incomplete transmission, denoted by $p_{k,i}$,
is $C_{(t-i-1)L-k+i}^{i}P_{nt}^{(t-i-1)L-k}P_t^i$. Furthermore, the
probability of the given node having $j$ successful complete
transmissions on condition that there are $i$ complete
transmissions, denoted by $p_{i,j}$, is $C_i^j (P_s/P_t)^j
(P_o/P_t)^{i-j}$. Finally, we have $\tE e^{\theta I(s,s+t)}$ is
upper-bounded by
\begin{eqnarray}\label{eq:EI}
P_t\cdot\sum_{k=1}^{L-1}\sum_{i=0}^{t-2}\sum_{j=0}^{i}p_{k,i}
p_{i,j} e^{\theta(t-j)} + \sum_{i=0}^{t-1}\sum_{j=0}^{i} p_{0,i}
p_{i,j} e^{\theta(t-j)}.
\end{eqnarray}
The first term is for the case that the last transmission is
incomplete and the second term is for the case that the last
transmission is complete.

In general, we do not have an analytical form of Eq.~(\ref{eq:EI}),
so we resort to numerical methods and use Algorithm 1 to obtain
$\sigma_{I}(\theta)$ and $\rho_{I}(\theta)$ (see Appendix A-1). This
algorithm is immediately inspired from
Definition~\ref{def:theta_uc}. Then we can use Eq.~(\ref{eq:I_ac})
and (\ref{eq:S_sc}) to obtain the node's weak stochastic service
curve.

\begin{figure}[htb]
\begin{center}
\begin{tabular}{|l|l|}
\hline
\textbf{Scenario 1:}\\
\ 10 identical nodes sending packets to one access point\\
\ The payload of a DATA packet is $256$ bytes\\
\hline
\end{tabular}
\end{center}
\caption{Parameters of Scenario 1}\label{fig:scen1}
\end{figure}

\begin{figure}[htb]
  \centering
  \includegraphics[width=0.6\textwidth]{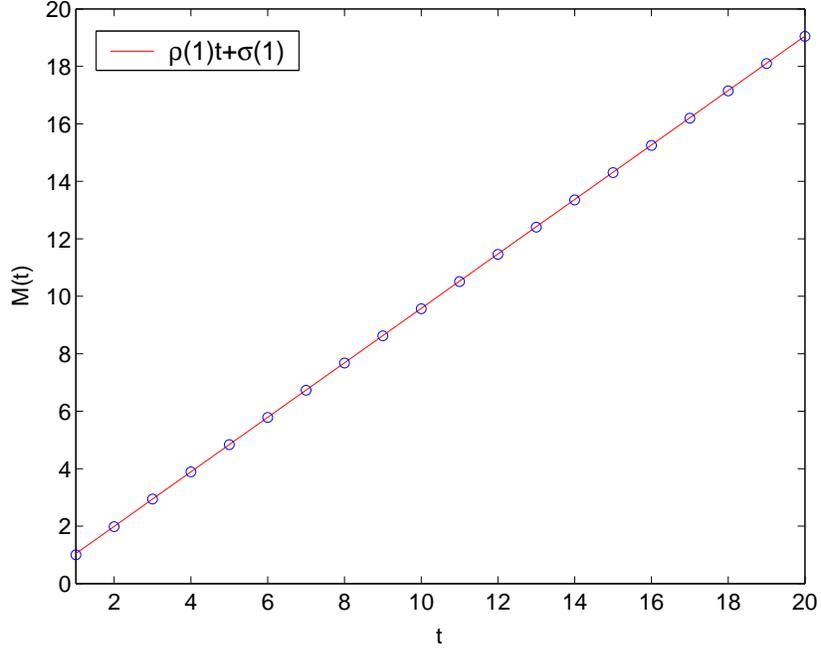}\\
  \caption{Scenario 1: the $(\sigma(1),\rho(1))$-upper constrained curve of a node's $M(t)$}\label{fig:ex_Mt}
\end{figure}

We illustrate the above calculations in Scenario 1 (see
Fig.~\ref{fig:scen1}). From Eq.~(\ref{eq:tau_gamma}) and
(\ref{eq:gamma_tau}), $\tau = 0.037$ and $\gamma = 0.293$. Thus,
$P_{nt}=0.680$, $P_{t}=0.320$, $P_s = 0.027$ and $P_o = 0.293$. Let
$M(t)$ be the value of Eq.~(\ref{eq:EI}). Fig.~\ref{fig:ex_Mt} shows
$M(t)$ and its $\sigma(\theta), (\rho(\theta))$-upper constrained
curve when $\theta=1$. We have $\rho(1)=0.948$ and
$\sigma(1)=0.096$, as calculated by Algorithm 1.

\begin{figure}[htb]
  \centering
  \includegraphics[width=0.6\textwidth]{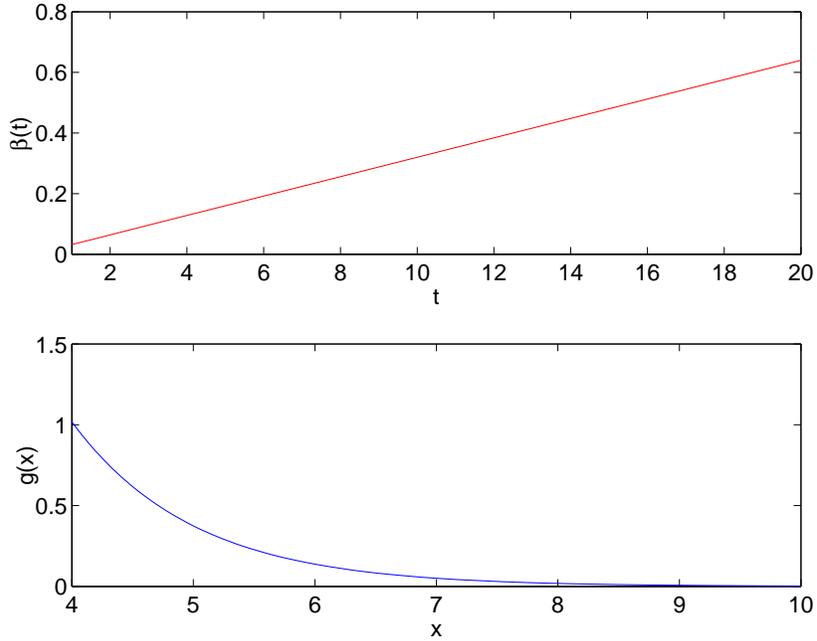}\\
  \caption{Scenario 1: a node's Weak stochastic service curve}\label{fig:ex_sc}
\end{figure}

Furthermore, from Eq.~(\ref{eq:I_ac}) and (\ref{eq:S_sc}) we have $S
\sim_{ws} <g, \beta>$ where
\begin{eqnarray*}
\beta(t) &=& (1-r_I)\cdot t\\
g(x) &=& \frac{e^{0.096}}{1-e^{0.948-r_I}}\cdot e^{-x},
\end{eqnarray*}
for any $1 > r_I > 0.948$. Fig.~\ref{fig:ex_sc} plots $\beta(t)$ and
$g(t)$ when $r_I=0.968$.

\subsection{{\bf Stability Condition}}\label{sec:802_11_stab}

Before calculating the bounds, we first derive the stability
condition. Since $\hat{\beta}(t)=t$, the channel capacity $c = 1$
packet per slot. Because $1-a_I$ is actually the percentage of the
node's successful transmission time, by Proposition~\ref{prop:stab},
the \emph{stability condition of 802.11} is
\begin{eqnarray}\label{eq:stab_802_11}
a_A < 1-a_I = \frac{P_s \cdot L}{P_{nt} + P_t \cdot L}.
\end{eqnarray}

\subsection{{\bf Backlog and Delay Bounds}}\label{sec:802_11_bounds}

We can immediately calculate backlog bounds by applying
Eq.~(\ref{eq:A_ac})-(\ref{eq:S_sc}) into
Theorem~\ref{theo:backlog_delay}. The only technical issue is to
select proper $\theta_1$ and $\theta_2$ a obtain tight bounds.
Clearly, $f(x)$ depends on $\theta_1$ and $r_A$, and $g(x)$ depends
on $\theta_2$ and $r_I$. According to Eq.~(\ref{eq:stab_1}), we have
$r_A \leq 1 - r_I$ and $P\{B(t)
> x\} \leq f\otimes g(x)$. In addition, $r_A$ ($r_I$) should be set as large as
possible because $f(x)$ ($g(x)$) decreases with $r_A$ ($r_I$).
Considering the above conditions, we have
\begin{eqnarray}\label{eq:my_backlog_bound}
&&P\{B(t) > x\} \leq \min[ f\otimes g(x)]\nonumber\\
&&\textnormal{subject to}\nonumber\\
&&r_A > \rho_A(\theta_1), r_I > \rho_I(\theta_2)\ \textnormal{and}\
r_A + r_I = 1.
\end{eqnarray}
In general, we do not have an analytical solution of $\min[ f\otimes
g(x)]$ and we resort to numerical methods and use Algorithm 2 to get
a near-optimal solution (see Appendix A-2).

As noticed recently by researchers of network calculus, the delay
bound in Theorem~\ref{theo:backlog_delay} often returns trivial
results. In our model in Section~\ref{sec:model}, it is easy to see
that $P\{D(t)
> x\} \leq f\otimes g(0) = 1$. We propose the following way to estimate delay
bound. Little's law states that the average number of customers in a
queueing system is equal to the average arrival rate of customers to
that system, times the average time spent in that
system\cite{queueing_book}. Let the average arrival rate is
$\lambda$. Assume the system can reach \emph{steady state} when
$t\rightarrow\infty$. Then we have the average backlog is
$\lim_{t\rightarrow\infty}\tE B(t)$ and the average delay of each
packet is greater than or equal to $\lim_{t\rightarrow\infty}\tE
D(t)$ (by its definition in Eq.~(\ref{eq:delay}), $D(t)$ can be less
than the delay of the bottom-of-line packet at $t$). Therefore, by
Little's law, we have
\begin{eqnarray}\label{eq:Little}
\lim_{t\rightarrow\infty}\tE D(t) \leq
\frac{\lim_{t\rightarrow\infty}\tE B(t)}{\lambda}.
\end{eqnarray}
Finally, we apply Markov's inequality to the above equation and we
have
\begin{eqnarray}\label{eq:my_delay_bound}
P\{\lim_{t\rightarrow\infty}D(t) \geq x\} \leq
\frac{\lim_{t\rightarrow\infty}\tE B(t)}{\lambda x}.
\end{eqnarray}
Besides, according to Eq.~(\ref{eq:EB}), $\tE B(t) \leq
\sum_{i=0}^\infty{P\{B(t)
> i\}\cdot(i+1)}$. And we can use Eq.~(\ref{eq:my_backlog_bound}) to bound $P\{B(t) >
i\}$. Note that Eq.~(\ref{eq:my_delay_bound}) is derived when
$t\rightarrow \infty$. In practice, we can use this result to
estimate delay bound when $t$ is sufficiently large.

\section{{\bf Performance Evaluation}}\label{sec:simulation}

In this section, we use ns-2 simulations to verify our derived
backlog and delay bounds for Poisson and constant bit rate (CBR)
traffic arrivals. We carry out all experiments for Scenario 1
(Fig.~\ref{fig:scen1}). Each simulation duration is 100 seconds
(\emph{s}) which is long enough to let a node transmit thousands of
packets. Each data point (e.g. $P\{B(t)>x\}$) is calculated over 100
independent simulations.

\subsection{{\bf Poisson Traffic}}

Let $\lambda$ be the average traffic rate (packets/slot). In this
case, we have $a_A = \lambda$ (see Definition~\ref{def:ER}). For
Poisson traffic, we have
\begin{eqnarray}
Ee^{\theta A(s,s+t)} = \sum_{i=0}^\infty \frac{(\lambda t)^i}{i!}
e^{-\lambda t} e^{\theta i} = e^{\lambda t(e^\theta-1)},
\end{eqnarray}
where $\frac{(\lambda t)^i}{i!} e^{-\lambda t}$ is the probability
of $i$ packets arriving within $(s, s+t]$. From the above equation,
Poisson traffic is $(0, \frac{\lambda(e^\theta-1)}{\theta})$-upper
constrained and we can obtain its arrival curve by
Eq.~(\ref{eq:A_ac}).

\begin{figure}[htb]
  \centering
  \includegraphics[width=0.6\textwidth]{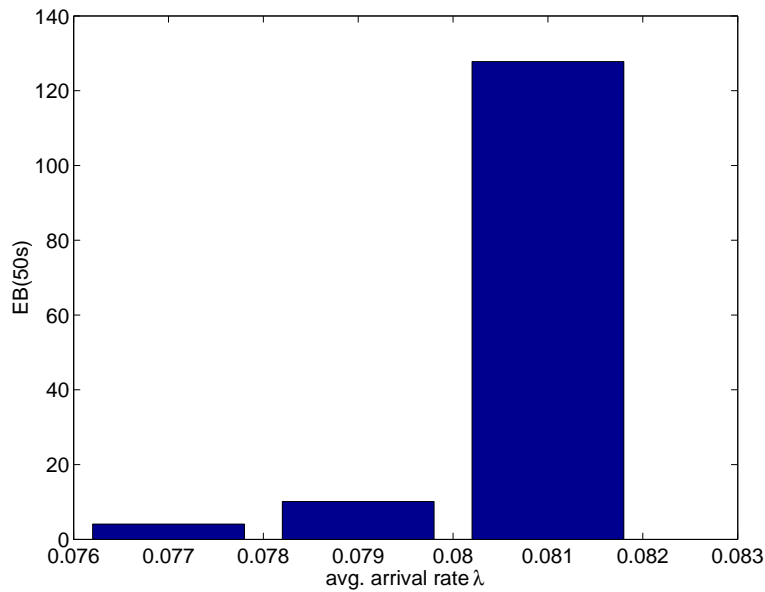}\\
  \caption{$\tE B(t)$ ($t=50$s) when $\lambda=0.077, 0.079, 0.081$}\label{fig:plot_10_stable_Poisson}
\end{figure}

From Eq.~(\ref{eq:stab_802_11}), backlogs are stable when $\lambda <
0.079$ (packet per slot). In Fig.~\ref{fig:plot_10_stable_Poisson},
we plot the average backlog $\tE[B(t)]$ at $t=50s$ and $\lambda =
0.077$, $0.079$ and $0.081$ in ns-2 simulations. We observe that
there is sudden jump when $\lambda = 0.081$, indicating the critical
point of stability is indeed around $0.079$. This figure also
indicates the accuracy of the 802.11 model in
Eq.~(\ref{eq:tau_gamma}) and Eq.~(\ref{eq:gamma_tau}).

\textbf{Experiment 1}(Scenario 1 with low Poisson traffic load)  We
set $\lambda = 0.04$ to simulate low traffic load.

\begin{figure}[htb]
  \centering
  \includegraphics[width=0.6\textwidth]{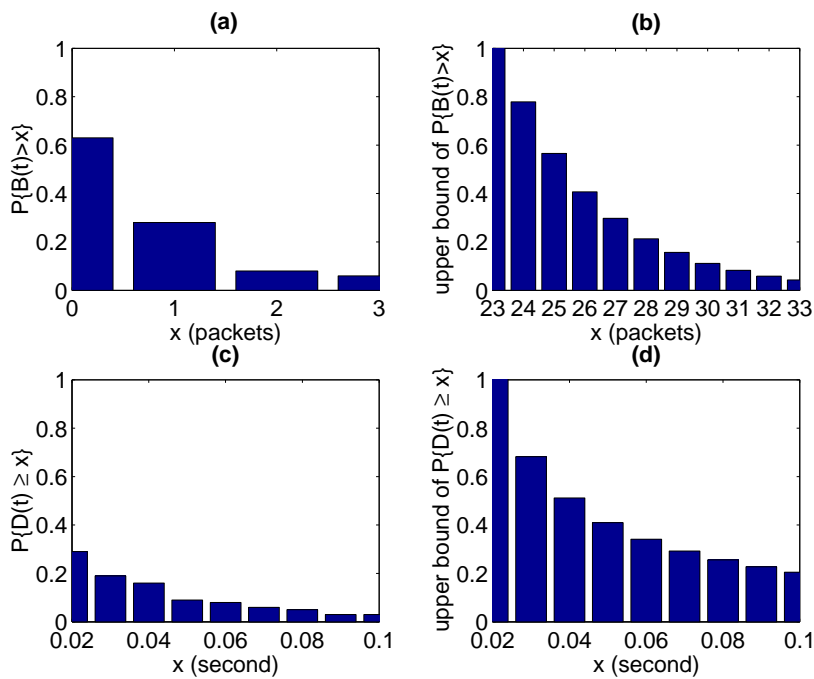}\\
  \caption{Experiment 1: When $t = 50s$\ (a) $P\{B(t)>x\}$ (b) upper bound of $P\{B(t)>x\}$ (c) $P\{D(t)\geq x\}$ (d) upper bound of $P\{D(t)\geq x\}$}\label{fig:plot_10_Poisson_low}
\end{figure}

Fig.~\ref{fig:plot_10_Poisson_low}(a) shows $P\{B(t)>x\}$ in ns-2
simulations and Fig.~\ref{fig:plot_10_Poisson_low}(b) shows the
upper bound of $P\{B(t)>x\}$ calculated by
Eq.~(\ref{eq:my_backlog_bound}). Note that stochastic network
calculus gives very loose upper bounds. There may be two reasons.
One is that we use the worst-case analysis in deriving the weak
stochastic service curve of 802.11. The other is that there are many
relaxations in proving the theorems of stochastic network
calculus\cite{snetcal_jiang}. For example, relaxations are used in
deriving $f(x)$ in Theorem~\ref{theo:ac_theta}
\cite{snetcal_book_jiang}, and this theorem is popularly used in
deriving arrival curves and service curves (see
Section~\ref{sec:model}). The first reason may not be the key reason
because we will see in Experiment 2 the bound is even looser when we
increase arrival rate and make the channel near saturated. The
second reason seems to be the key reason. We will see in Experiment
3 that backlog bounds improve substantially for CBR traffic where we
are able to derive the arrival curve by hand without using
Theorem~\ref{theo:ac_theta}. This indicates that refinements are
needed in stochastic network calculus so as to tighten the bounds.
Moreover, we found that backlog bounds are sensitive to adjusting
parameters (i.e., $\theta_1$, $\theta_2$, $r_A$ and $r_I$). So it is
necessary to use Algorithm 2 to minimize the bounds.

We also conduct simulations to verify delay bounds at $t=50s$. Since
the backlog bounds are too loose, in order to avoid trivial
validation, we use $\tE B(t)$ in ns-2 simulations to validate
Eq.~(\ref{eq:Little}) and Eq.~(\ref{eq:my_delay_bound}) (assume that
$t=50$s is sufficiently large so that we can apply these equations).
Actually, we get $\tE D(t) \leq 0.0205s$ by Eq.~(\ref{eq:Little}),
which tightly bounds $\tE D(t) = 0.0186s$ in ns-2 simulations.
Fig.~\ref{fig:plot_10_Poisson_low}(c) shows $P\{D(t)\geq x\}$ in
ns-2 simulations and Fig.~\ref{fig:plot_10_Poisson_low}(d) shows the
upper bound of $P\{D(t)\geq x\}$ calculated by
Eq.~(\ref{eq:my_delay_bound}). Clearly, $P\{D(t)\geq x\}$ is
upper-bounded by Eq.~(\ref{eq:my_delay_bound}).

\textbf{Experiment 2}(Scenario 1 with high Poisson traffic load)  We
set $\lambda = 0.07$ to simulate high traffic load.

\begin{figure}[htb]
  \centering
  \includegraphics[width=0.6\textwidth]{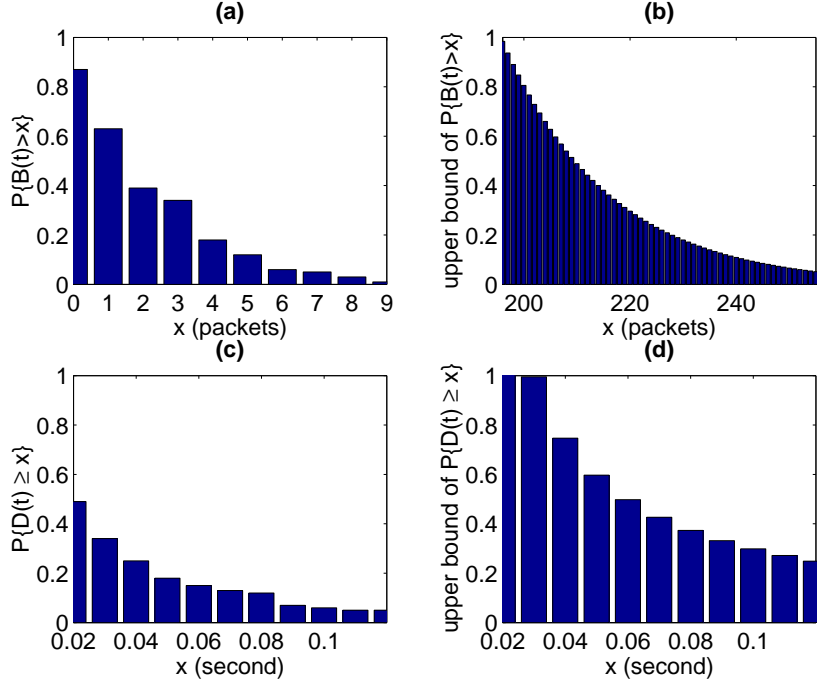}\\
  \caption{Experiment 2: When $t = 50s$\ (a) $P\{B(t)>x\}$ (b) upper bound of $P\{B(t)>x\}$ (c) $P\{D(t)\geq x\}$ (d) upper bound of $P\{D(t)\geq x\}$}\label{fig:plot_10_Poisson_high}
\end{figure}

Fig.~\ref{fig:plot_10_Poisson_high}(a) shows $P\{B(t)>x\}$ in ns-2
simulations and Fig.~\ref{fig:plot_10_Poisson_high}(b) shows the
upper bound of $P\{B(t)>x\}$ calculated by
Eq.~(\ref{eq:my_backlog_bound}). In this case, $\theta_1$
($\theta_2$) is much smaller than that of Experiment 1 so as to
satisfy the constraint in Eq.~(\ref{eq:my_backlog_bound}), resulting
in looser $f(x)$ and $g(x)$. Therefore, stochastic network calculus
gives further loose backlog bounds.

We also conduct simulations to verify delay bounds at $t=50s$. Since
the backlog bounds are too loose, in order to avoid trivial
validation, we use $\tE B(t)$ in ns-2 simulations to validate
Eq.~(\ref{eq:Little}) and Eq.~(\ref{eq:my_delay_bound}) (assume that
$t=50$s is sufficiently large so that we can apply these equations).
Actually, we get $\tE D(t) \leq 0.0299s$ by Eq.~(\ref{eq:Little}),
which tightly bounds $\tE D(t) = 0.0296s$ in ns-2 simulations.
Fig.~\ref{fig:plot_10_Poisson_high}(c) shows $P\{D(t)\geq x\}$ in
ns-2 simulations and Fig.~\ref{fig:plot_10_Poisson_high}(d) shows
the upper bound of $P\{D(t)\geq x\}$ calculated by
Eq.~(\ref{eq:my_delay_bound}). Clearly, $P\{D(t)\geq x\}$ is
upper-bounded by Eq.~(\ref{eq:my_delay_bound}).

\subsection{{\bf CBR Traffic}}

Let $\lambda$ be the average traffic rate (packets/slot). In this
case, we have $a_A = \lambda$. It is easy to see that $\sup_{0 \leq
s \leq t}[A(s, t) - \lambda \cdot (t-s)] < 1$ for all $t$ because
packets arrive one by one in a constant time interval. Thus, we have
$A \sim_{vb} <f, \alpha>$ where
\begin{eqnarray}
\alpha(t) &=& \lambda t \nonumber\\
f(x) &=& \left\{ \begin{array}{ll}
0, & \ x \geq 1\\
1, & \ x < 1
\end{array} \right.
\end{eqnarray}

\begin{figure}[htb]
  \centering
  \includegraphics[width=0.6\textwidth]{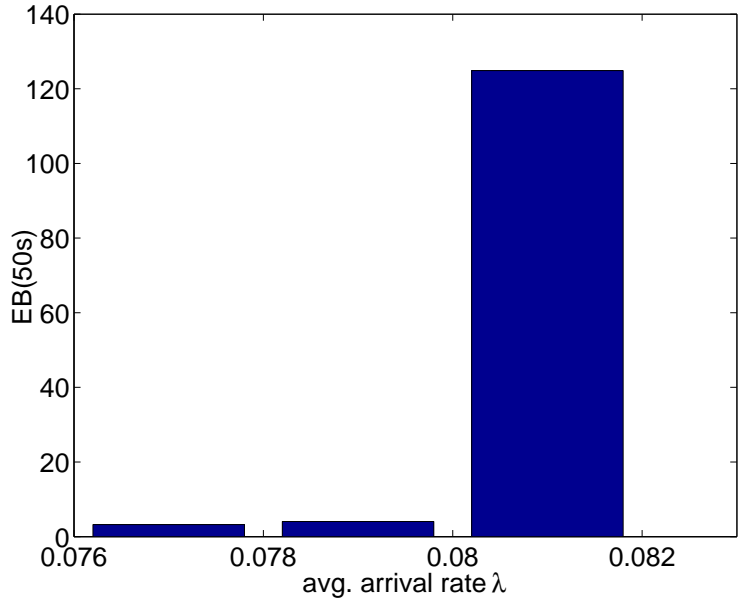}\\
  \caption{$\tE B(t)$ ($t=50$s) when $\lambda=0.077, 0.079, 0.081$}\label{fig:plot_10_stable_CBR}
\end{figure}

Again, from Eq.~(\ref{eq:stab_802_11}), the stability condition is
$\lambda < 0.079$ (packet per slot). In
Fig.~\ref{fig:plot_10_stable_CBR}, we plot the average backlog
$\tE[B(t)]$ at $t=50s$ for $\lambda = 0.077$, $0.079$ and $0.081$ in
ns-2 simulations. We observe that there is sudden jump when $\lambda
= 0.081$, indicating the critical point of stability is indeed
around $0.079$. This figure also indicates the accuracy of the
802.11 model in Eq.~(\ref{eq:tau_gamma})and(\ref{eq:gamma_tau}).

\textbf{Experiment 3}(Scenario 1 with low CBR traffic load)  We set
$\lambda = 0.04$ to simulate low traffic load.

\begin{figure}[htb]
  \centering
  \includegraphics[width=0.6\textwidth]{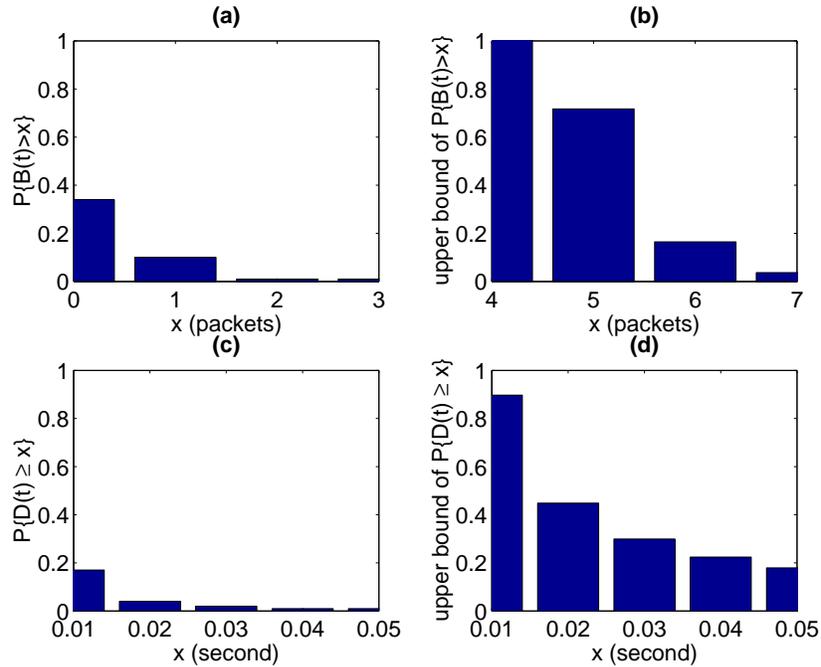}\\
  \caption{Experiment 3: When $t = 50s$\ (a) $P\{B(t)>x\}$ (b) upper bound of $P\{B(t)>x\}$ (c) $P\{D(t)\geq x\}$ (d) upper bound of $P\{D(t)\geq x\}$}\label{fig:plot_10_CBR_low}
\end{figure}

Fig.~\ref{fig:plot_10_CBR_low}(a) shows $P\{B(t)>x\}$ in ns-2
simulations and Fig.~\ref{fig:plot_10_CBR_low}(b) shows upper bound
of $P\{B(t)>x\}$ calculated by Eq.~(\ref{eq:my_backlog_bound}). The
backlog bounds are much tighter in CBR traffic than those in Poisson
traffic (see Experiment 1). The main reason is that we can derive a
tight $f(x)$ by hand instead of by Theorem~\ref{theo:ac_theta}.

We also conduct simulations to verify delay bounds at $t=50s$. Since
the backlog bounds are still loose, in order to avoid trivial
validation, we use $\tE B(t)$ in ns-2 simulations to validate
Eq.~(\ref{eq:Little}) and Eq.~(\ref{eq:my_delay_bound}) (assume that
$t=50$s is sufficiently large so that we can apply these equations).
Actually, we get $\tE D(t) \leq 0.0090s$ by Eq.~(\ref{eq:Little}),
which tightly bounds $\tE D(t) = 0.0089s$ in ns-2 simulations.
Fig.~\ref{fig:plot_10_CBR_low}(c) shows $P\{D(t)\geq x\}$ in ns-2
simulations and Fig.~\ref{fig:plot_10_CBR_low}(d) shows the upper
bound of $P\{D(t)\geq x\}$ calculated by
Eq.~(\ref{eq:my_delay_bound}). Clearly, $P\{D(t)\geq x\}$ is
upper-bounded by Eq.~(\ref{eq:my_delay_bound}).

\textbf{Experiment 4}(Scenario 1 with high CBR traffic load)  We set
$\lambda = 0.07$ to simulate high traffic load.

\begin{figure}[htb]
  \centering
  \includegraphics[width=0.6\textwidth]{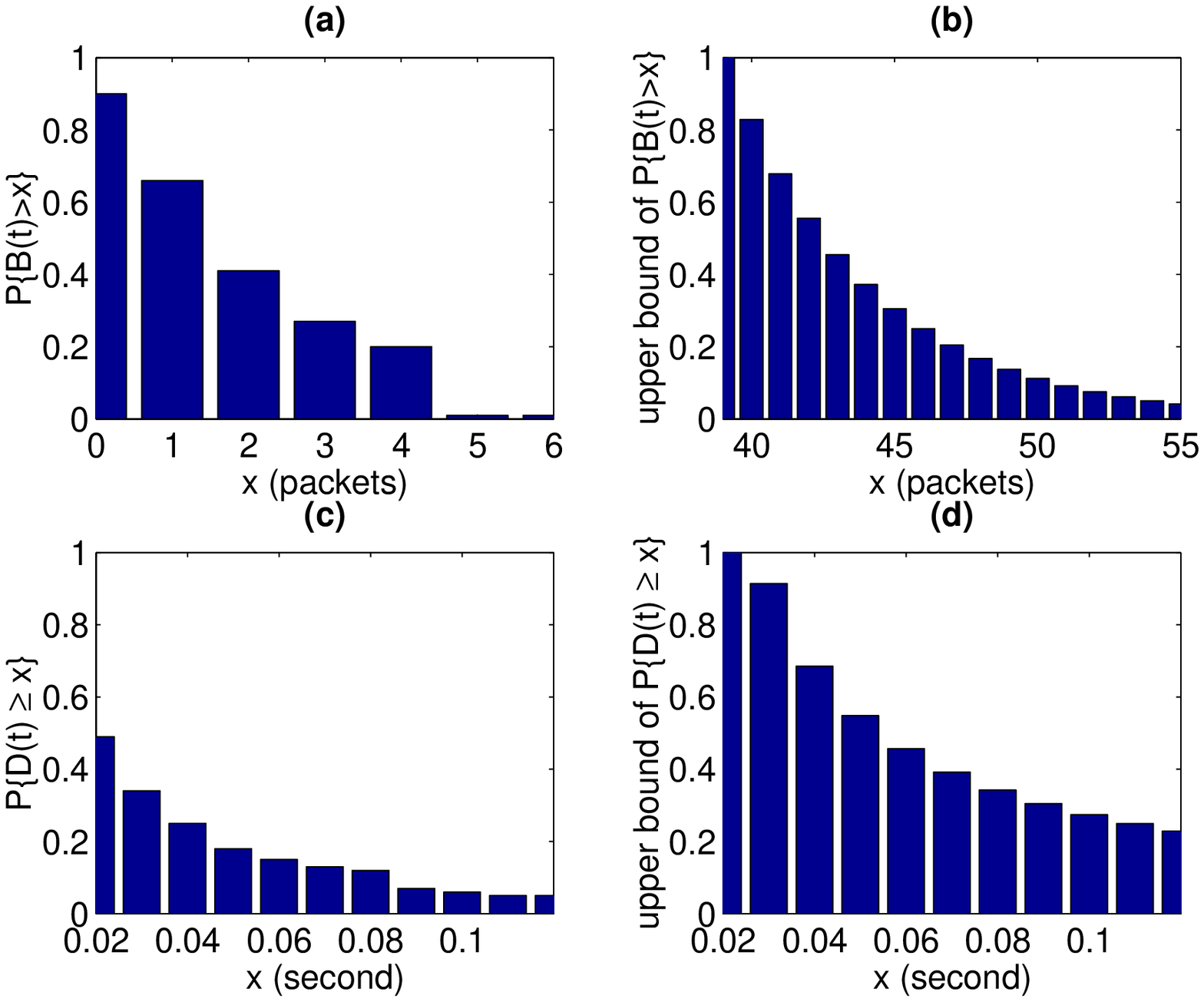}\\
  \caption{Experiment 4: When $t = 50s$\ (a) $P\{B(t)>x\}$ (b) upper bound of $P\{B(t)>x\}$ (c) $P\{D(t)\geq x\}$ (d) upper bound of $P\{D(t)\geq x\}$}\label{fig:plot_10_CBR_high}
\end{figure}

Fig.~\ref{fig:plot_10_CBR_high}(a) shows $P\{B(t)>x\}$ in ns-2
simulations and Fig.~\ref{fig:plot_10_CBR_high}(b) shows the upper
bound of $P\{B(t)>x\}$ calculated by
Eq.~(\ref{eq:my_backlog_bound}). The backlog bounds are much tighter
in CBR traffic than those in Poisson traffic (see Experiment 2)
because $f(x)$ is tight here.

We also conduct simulations to verify delay bounds at $t=50s$. Since
the backlog bounds are still loose, in order to avoid trivial
validation, we use $\tE B(t)$ in ns-2 simulations to validate
Eq.~(\ref{eq:Little}) and Eq.~(\ref{eq:my_delay_bound}) (assume that
$t=50$s is sufficiently large so that we can apply these equations).
Actually, we get $\tE D(t) \leq 0.0274s$ by Eq.~(\ref{eq:Little}),
which is close to (although does not bound) $\tE D(t) = 0.0296s$ in
ns-2 simulations. Fig.~\ref{fig:plot_10_CBR_high}(c) shows
$P\{D(t)\geq x\}$ in ns-2 simulations and
Fig.~\ref{fig:plot_10_CBR_high}(d) shows the upper bound of
$P\{D(t)\geq x\}$ calculated by Eq.~(\ref{eq:my_delay_bound}).
Clearly, $P\{D(t)\geq x\}$ is upper-bounded by
Eq.~(\ref{eq:my_delay_bound}).

To sum up, the current version of stochastic network calculus often
derives loose bounds when compared with simulations, especially in
the case of high traffic load. Therefore, stochastic network
calculus may not be effective in practice.

\section{{\bf Related Work}}\label{sec:related}
In this section, we first present relate work on stochastic network
calculus and then on the performance analysis of 802.11.

The increasing demand on transmitting multimedia and other real time
applications over the Internet has motivated the study of quality of
service guarantees. Towards it, stochastic network calculus, the
probabilistic version of the deterministic network calculus
\cite{netcal_cruz_1}\cite{netcal_cruz_2}\cite{netcal_boudec}\cite{netcal_chang},
has been recognized by researchers as a promising step. During its
development, traffic-amount-centric (t.a.c) stochastic arrival curve
is proposed in \cite{snetcal_li}, virtual-backlog-centric (v.b.c)
stochastic arrival curve is proposed in \cite{snetcal_jiang_o1} and
maximum-backlog-centric (m.b.c) stochastic arrival curve is proposed
in \cite{snetcal_jiang}. Weak stochastic service curve is proposed
in \cite{snetcal_cruz}\cite{snetcal_liu} and stochastic service
curve is proposed in \cite{snetcal_jiang_o2}. In
\cite{snetcal_jiang}, Jiang showed that only the combination of
m.b.c stochastic arrival curve and stochastic service curve has all
five basic properties required by a network calculus (i.e.,
superposition, concatenation, output characterization, per-flow
service, service guarantees) and the other combinations only have
parts of these properties. Jiang also proposed the concept of
stochastic strict server to facilitate calculation of stochastic
service curve. Moreover, he presented independent case analysis to
obtain tighter performance bounds for the case that flows and
servers are independent. However, there are a few bugs in his
results recently found by researchers of network calculus, such as
the trivial delay bound in Theorem~3.5\cite{snetcal_jiang} and
Theorem~5.1. Therefore, we adopt v.b.c stochastic arrival curve and
weak stochastic service curve in our study since we only consider
backlog and delay bounds (i.e., service guarantee), ignoring the
other properties.

There have been some applications of stochastic network calculus. In
\cite{app_MBAC}, Jiang et al. analyzed a dynamic priority
measurement-based admission control (MBAC) scheme based on
stochastic network calculus. In \cite{app_SLA}, Liu et al. applied
stochastic network calculus to studying the conformance
deterioration problem in networks with service level agreements. In
\cite{app_LRD_GPS}, based on stochastic network calculus, X. Yu et
al. developed several upper bounds on the queue length distribution
of Generalized Processor Sharing (GPS) scheduling discipline with
long range dependent (LRD) traffic. They also extended the GPS
results to a packet-based GPS (PGPS) system. Finally, Agharebparast
et al. modeled the behavior of a single wireless link using
stochastic network calculus \cite{app_wireless}. However, little
effort has been made on applying stochastic network calculus to
multi-access communication systems such as 802.11.

Existing work on the performance of 802.11 has focused primarily on
its throughput and capacity. In \cite{Bianchi_802_11}, Bianchi
proposed a Markov chain throughput model of 802.11. In
\cite{Kumar_802_11}, Kumar et al. proposed a probability throughput
model which is simpler than Bianchi's model. In our paper, we adopt
Kumar's model to derive the service curve of 802.11. There are also
some work on queueing analysis of 802.11. In \cite{Zhai_802_11},
Zhai et al. assumed Poisson traffic arrival and proposed an M/G/1
queueing model of 802.11. More generally, Tickoo proposed a G/G/1
queueing model of 802.11\cite{Tickoo1_802_11}\cite{Tickoo2_802_11}.
To our best knowledge, we are the first to model the queueing
process of 802.11 based on stochastic network calculus.

\section{{\bf Conclusion and Future Work}}\label{sec:conclusion}
In this paper, we presented a stochastic network calculus model of
802.11. From stochastic network calculus, we first derived the
general stability condition of a wireless node. Then we derived the
stochastic service curve and the specific stability condition of an
802.11 node based on an existing model of 802.11. Thus, we obtained
the backlog and delay bounds of the node by using the corresponding
theorem of stochastic network calculus. Finally, we carried out ns-2
simulations to verify these bounds.

There are some open problems for future work. First, we derived the
service curve based on an existing 802.11 model. Thus, the accuracy
of the service curve depends on the accuracy of the model. An open
question may be whether we can derive the service curve of 802.11
without using any existing models. Second, we assumed the worst-case
condition (i.e., saturate condition) in our analysis. Can we remove
this conservative assumption? Besides, under the worst-case
assumption, we can assume flows and servers are independent and
perform independent case analysis obtaining tighter backlog and
delay bounds. This is also one of our future work. Third, we observe
that the derived bounds are loose when compared with ns-2
simulations, calling for further improvements in the current version
of stochastic network calculus.

\begin{center}
    {\bf Appendix A-1: Algorithm 1 \emph{(Numerical Calculation of
$\sigma_{I}(\theta)$ and $\rho_{I}(\theta)$)}}
\end{center}

\begin{enumerate}
\item Let $M(t)= \sup_{s\geq 0}\{\frac{1}{\theta}\log{\tE e^{\theta I(s,s+t)}}\}$. Obviously, $M(t)$ is an increasing function of $t$
with $M(0)=0$. We define axes $t$ and axes $t_\bot$ (vertical to
$t$) on a plane, and plot $M(t)$ on it. We define the slope of
$M(t)$, $s(t) = M(t)-M(t-1)$.
\item We calculate $s(t)$ for $t = 1, 2, 3,...$ until it converges at some
$t^*$, i.e., $(1-\epsilon) \cdot s(t^*-1) \leq s(t^*) \leq
(1+\epsilon) \cdot s(t^*-1)$ where $\epsilon$ is a small number,
e.g. $10^{-5}$.
\item We draw a straight line $l(t)$ with the slope $s(t^*)$ crossing the point $\big(t^*,
M(t^*)\big)$. Obviously, the line crosses the point $\big(0,
M(t^*)-s(t^*)t^*\big)$. The maximum displacement between $M(t)$ and
$l(t)$ (in the direction of $t_\bot$), $v_m=\max_{0\leq t\leq
t^*}\{M(t)-l(t)\}$. We shift $l(t)$ by $v_m$ in the direction of
$t_\bot$ and get $\tilde {l}(t)$. Clearly, $\tilde {l}(t)$
upperbounds $M(t)$. In other words, we have
$\rho_{I}(\theta)=s(t^*)$ and $\sigma_{I}(\theta)
=M(t^*)-s(t^*)t^*+v_m$.
\end{enumerate}

\begin{center}
    {\bf Appendix A-2: Algorithm 2 \emph{(Numerical Calculation of
Eq.~(\ref{eq:my_backlog_bound}))}}
\end{center}

In each iteration, we generate a sample of $\theta_1$, $\theta_2$,
$r_A$ and $r_I$. If they satisfy the condition of
Eq.~(\ref{eq:my_backlog_bound}), we calculate $\min[f\otimes g(x)]$
for the current and past iterations until it converges. Sample
generations can use the interpolation or Monte Carlo method over
valid ranges of the variables.

\end{document}